\documentclass[11pt]{article}
\usepackage[margin=25mm]{geometry}
\usepackage{graphicx}
\usepackage{xcolor}
\usepackage{xurl}
\usepackage[T1]{fontenc}
\usepackage{lmodern}
\usepackage{textcomp}
\usepackage{amsmath}
\usepackage{amssymb}
\usepackage[round]{natbib}

\title{Statistical Analysis of Executability and Program Equivalence\\
 in Decompilation for IoT Vulnerability Detection\thanks{This is the
 authors' English translation of a paper (in Japanese) accepted for publication in the
 special issue ``Theory and Applications of Privacy Protection'' of
 Toukei Suri (Proceedings of the Institute of Statistical Mathematics).
 \copyright~The Institute of Statistical Mathematics.}}

\author{Minami Yoda$^{1}$ \and Jialong Li$^{2}$ \and Yasuyuki Tahara$^{3}$
 \and Yuichi Sei$^{3}$ \and Yutaka Matsuno$^{1}$}
\date{%
 \small
 $^{1}$College of Science and Technology, Nihon University, Chiba, Japan\\
 $^{2}$Waseda Institute for Advanced Study, Waseda University, Tokyo, Japan\\
 $^{3}$Graduate School of Informatics and Engineering,
 The University of Electro-Communications, Tokyo, Japan\\[1ex]
 \texttt{yoda.minami@nihon-u.ac.jp}}

\begin{document}
\maketitle

\begin{abstract}
Internet of Things (IoT) devices handle sensitive privacy-related information such as user audio, video, and authentication data, making it indispensable to detect vulnerabilities inherent in their firmware. Decompilation, one of the key detection techniques, has recently attracted attention because the use of Large Language Models (LLMs) enables high readability and high recompilation success rates.

Decompilation is required to faithfully reconstruct the original program. However, because LLM outputs depend on probabilistic token prediction learned from training data, they tend to prioritize syntactic correctness and thus generate plausible-looking code that is semantically different from the original binary. In particular, vulnerabilities arise in details that are easily lost through the probabilistic generation of LLMs, such as error-handling flows and insufficient boundary checks, so the generated results may differ from the original. Existing evaluation metrics focus mainly on passing test cases and therefore cannot sufficiently identify code whose internal structure has been altered even though it appears behaviorally valid. Consequently, a new evaluation metric that quantifies the internal structure of generated results from multiple perspectives is needed.

In this study, we propose a nine-dimensional quality evaluation metric consisting of three major categories---structural, behavioral, and semantic similarity---and quantify the degree of restoration of the generated results. In our experiments, targeting 318 programs from OpenWrt, an open-source router platform adopted as the basis of many commercial routers, we generated 19,625 decompilation results using five decompilation methods (one rule-based and four LLM-based) and analyzed them statistically. The results show that the recompilation-success group achieved significantly higher overall scores than the failure group, with an effect size of Cohen's $d=0.92$.
In particular, behavioral similarity exhibited an effect size of $d = 0.96$ and structural similarity $d = 0.69$, demonstrating that these metrics are important predictors of decompilation quality.
This study provides a statistical evaluation foundation for quantifying implementation defects in IoT devices and presents a framework that also generalizes to quality evaluation of black-box generative models.
\end{abstract}

\noindent\textbf{Keywords:} Decompilation, Large Language Models, Program Equivalence, IoT Security, Statistical Evaluation, Binary Analysis

\section{Introduction}
The number of deployed Internet of Things (IoT) devices is predicted to reach 52.2 billion in 2026 and 58.3 billion in 2027; because identical models are deployed in large numbers, the impact of a single vulnerability is enormous \citep{soumu2025}.
In particular, IoT devices handle sensitive privacy-related information such as user audio, video, and authentication data, making it indispensable to detect vulnerabilities inherent in their firmware.
Vulnerabilities that actually led to privacy information leakage include the leakage of authentication credentials and eavesdropping on audio and video in web cameras (CVE-2021-28372), and the leakage of the shared key used in Wi-Fi bridge communication (CVE-2019-17098). These issues were discovered by decompiling and analyzing the firmware of the affected devices.

Decompilation is a technique that converts a binary into assembly and further restores it to C code. In IoT vulnerability analysis, decompilation is used to detect vulnerabilities. The goal of decompilation is to produce code that exhibits functionally identical behavior to the original program \citep{Cifuentes1995}, and various methods have been proposed to achieve this. Conventional decompilation technology has been dominated by rule-based methods such as Ghidra \citep{NSA2019}, but in recent years, methods leveraging Large Language Models (LLMs) have been proposed \citep{wong2023decgpt,DLiFT2025,Tan2024}.

Methods leveraging LLMs have made it possible to recompile the restored programs, improving re-executability. However, because LLM outputs are the result of probabilistic token prediction based on training data, they generate superficially plausible code; even if recompilation succeeds, the content may differ from the original program. Because the generation process of LLMs is opaque in this way, quantitative metrics are used to evaluate the quality of their outputs.

Prior studies have adopted behavioral and semantic similarity of programs as evaluation metrics. For example, MSSC in DeGPT \citep{wong2023decgpt} evaluates the agreement of return values and function calls, and D-Score in D-LiFT \citep{DLiFT2025} evaluates correctness and readability in an integrated manner. However,
existing evaluation metrics focus mainly on passing test cases and cannot sufficiently identify code whose internal structure has been altered even though it appears behaviorally valid. Moreover, vulnerability analysis of IoT devices requires the reproduction of structural details, such as a missing NULL check on the return value of fopen() or a missing free() in an error branch, so structural information such as control flow must also be evaluated.

Therefore, in this study we propose an evaluation metric consisting of three major categories---structural, behavioral, and semantic similarity---with nine dimensions in total. In addition to the behavioral and semantic similarity used in prior work, this metric introduces structural similarity. The purpose of structural similarity is to measure the degree of restoration of the control structures and data flows inherent in the binary.
In our experiments,
we statistically analyzed generation results targeting the open-source OpenWrt router, which is adopted as the basis of commercial routers. For 318 programs, using one rule-based method and four LLMs,
four types of prompts, and four compiler options, we generated 19,625 C programs and analyzed them statistically to verify the usefulness of the proposed metrics and the quality of the generated programs.
The contributions of this study are as follows.

\begin{enumerate}
\item We propose an integrated evaluation metric consisting of three major categories---structural similarity, behavioral similarity, and semantic similarity---with nine dimensions in total.

\item Based on the proposed metrics, we statistically analyze a large-scale dataset of 19,625 samples and identify the factors that predict decompilation quality.

\item We analyze cases in which recompilation fails despite high behavioral similarity, revealing that the cause lies in defects that are overlooked by behavioral evaluation but detectable by structural evaluation, thereby demonstrating the complementary relationship between the two.

The remainder of this paper is organized as follows. Section 2 reviews related work, Section 3 details the design of the proposed metrics, Section 4 presents the experimental setup and results, Section 5 provides discussion, and Section 6 concludes with future work.
\end{enumerate}

\section{Related Work}
\label{sec:background}

Vulnerability detection methods for IoT devices are broadly classified into static analysis, dynamic analysis, and methods combining both \citep{Feng2023IoTSurvey,IoTSecuritySurvey2024}. Static analysis includes methods based on firmware binary analysis and decompilation; it has the advantage of not requiring an execution environment and enabling large-scale analysis. Dynamic analysis, on the other hand, can directly observe runtime behavior but requires physical devices or the construction of an emulation environment.

This study focuses on static analysis, in particular on the quality evaluation of decompilation results. Privacy violations often stem from implementation-level defects, but if the quality of the analysis results obtained through decompilation is insufficient, their detection and assessment cannot be trusted either. Therefore, a framework for objectively evaluating the quality of decompilation results is also important for complementing privacy risk analysis.

\subsection{Rule-based Decompilation}
Conventional decompilation technology restores source code based on pattern matching and program analysis. Industry-standard tools such as IDA Pro \citep{IDAPro}, Ghidra \citep{NSA2019}, and the Hex-Rays Decompiler reconstruct control flow graphs and infer data types using static analysis. However, because the targets of function pointers and indirect jumps are determined at runtime, the control flow graph may be incomplete with static analysis alone.

Dynamic analysis frameworks such as angr identify these jump targets through symbolic execution and restore more complete control flow graphs \citep{angr}. However, these methods tend to prioritize code readability over generating recompilable and executable code. The limitations of conventional methods have also been confirmed quantitatively: Sirlan\v{c}i et al.\ analyzed more than one million outputs of major decompilers and reported that even syntactically correct code contains numerous semantic defects that produce execution results different from the original binary \citep{Sirlanci2025}.

\subsection{LLM-based Decompilation}
As a model dedicated to decompilation, SLaDe treats decompilation as a machine translation task using a Transformer and achieved a higher executable-program restoration rate than the rule-based method Ghidra \citep{SLaDe}. Because this approach restores code at the function level and requires a separate model for each architecture, generating executable files is difficult.

LLM Compiler, proposed by Meta, is a model built on Code Llama that targets translation between compiler intermediate representation (IR) and assembly, as well as code size optimization \citep{metallmcompiler2024}. This model was trained on a corpus of 546 billion tokens consisting of LLVM-IR and assembly code. It reproduces 77\% of the optimization candidates generated by autotuning and achieves 45\% accuracy (14\% exact match) in bidirectional translation between IR and assembly. Regarding BLEU scores, the FTD 7B model achieves 0.95 and the 13B model achieves 0.96, both showing high performance. However, the evaluation does not address the executability of the generated code.

LLM4Decompile \citep{Tan2024} fine-tuned DeepSeek Coder on ExeBench \citep{exebench}, a large-scale assembly dataset, and achieved performance surpassing GPT-4 and Ghidra in generating executable files. This dataset contains 7.2 million C functions compilable on x86 Linux. Although ExeBench also provides an ARM assembly dataset, the method of Tan et al.\ limits fine-tuning to x86 only.

Efforts to improve the readability of restored code are also advancing.
DeGPT \citep{wong2023decgpt} proposes a framework that uses LLMs to improve the readability of decompilation results. It divides the process into three roles---policy decision, revision proposal, and semantic verification---and performs variable renaming, comment insertion, and structure simplification step by step. For verifying semantic consistency, it uses MSSC (Micro Snippet Semantic Calculation), which compares changes in symbol values for random inputs. D-LiFT \citep{DLiFT2025} fine-tunes an LLM via reinforcement learning, aiming to improve readability while maintaining correctness. Its proposed D-Score first verifies syntactic and semantic correctness using a compiler and symbolic execution, and evaluates readability only for correct code.

However, these evaluation metrics share a common limitation. MSSC evaluates only the agreement of return values and external function calls, without considering the internal control flow or data structures of the program. D-Score includes correctness verification, but its readability evaluation relies on superficial indicators such as variable names and the presence of goto statements. Even in DecompileBench, the latest benchmark, the emphasis is placed on evaluating execution correctness, while the structural similarity of program control flow graphs and data dependencies is not evaluated \citep{DecompileBench2025}.
Under such evaluation, even if the program's execution results match the test cases, the reproduction of the internal structure cannot be guaranteed.

\subsection{Evaluation Metrics for Code Generation}
CodeBLEU is widely used as an evaluation metric for code generation, but its reliability has been questioned \citep{CodeBLEU2020}. Evtikhiev et al.\ investigated six major metrics including CodeBLEU and showed that their correlation with human evaluation is insufficient. They reported the paradoxical result that ChrF, a character-level comparison, is closer to human evaluation than CodeBLEU, which takes syntax into account \citep{Evtikhiev2023}. Quality concerns about existing benchmarks have also been pointed out \citep{BenchmarksMetrics2024}. In response, CodeScore proposes a method that uses LLMs to learn executability and related properties, improving the correlation with human evaluation by up to 58.87\% \citep{CodeScore2024}; however, these metrics target general code generation, and evaluation metrics specific to decompilation are needed.



\section{Design of the Evaluation Metrics}
\subsection{Design Policy}
The proposed metric consists of three major categories---structural similarity, behavioral similarity, and semantic similarity---each composed of three sub-metrics, for a total of nine dimensions.
Structural similarity evaluates whether the structure of the program is similar to the original binary, behavioral similarity evaluates whether the behavior of the program is equivalent to the original binary, and semantic similarity evaluates the semantic similarity of the program, such as variable names.

\subsection{Structural Similarity}
Structural similarity is a metric that evaluates whether the structure of the program is similar to the original binary. A binary retains function boundary information, control flow transitions, and the memory layout of data, all of which are fundamental structural information that decompilation should restore. In this study, we designed a metric consisting of three dimensions corresponding to this structural information: function signature completeness, CFG structural similarity, and data structure restoration.

\subsubsection{Function Signature Completeness}
Function signature completeness $S_{\text{sig}}$ evaluates the degree of agreement of function type information. It is defined as a weighted sum of the return type agreement $s_{\text{ret}}$, argument type agreement $s_{\text{arg\_type}}$, and argument count agreement $s_{\text{arg\_num}}$.
That is, $S_{\text{sig}} = 0.4 s_{\text{ret}} + 0.4 s_{\text{arg\_type}} + 0.2 s_{\text{arg\_num}}$, with a maximum score of 1.0.
Note that partial matches are permitted only for the return type, with an exact match scored as 0.4 and a partial match as 0.2.

\subsubsection{CFG Structural Similarity}
The Control Flow Graph (CFG) is a technique for representing the execution flow of a program as a directed graph and is one of the representative intermediate representations in program analysis \citep{cfg}.
Each node represents a basic block, and each edge represents a control transition. A basic block is a sequence of instructions executed sequentially, and control transitions include conditional branches, loops, and function calls. For example, an if statement branches from a condition node into two edges (true and false), and a while statement has a loop-back edge.

In this study, we extract the number of nodes, the number of edges, and the cyclomatic complexity from the CFG of each function and compute the CFG structural similarity based on them.
Cyclomatic complexity is a metric that quantifies the complexity of a program's control flow structure \citep{mccabe1976complexity}.
This metric represents the number of independent execution paths in a program, and its value increases with the number of conditional branches and loop structures.

Let $N, E, C$ denote the number of nodes, number of edges, and cyclomatic complexity of the original program, respectively, and $N', E', C'$ those of the decompiled program. Here, $x, y$ denote the two non-negative integers to be compared; in this study, we substitute $(x,y)=(N,N')$, $(E,E')$, and $(C,C')$ for the number of nodes, number of edges, and cyclomatic complexity, respectively, to compute the similarity. The similarity of two non-negative integers $x, y$ is defined as
\begin{equation}
s(x,y) = 1 - \frac{|x-y|}{\max(x,y)}
\end{equation}
where $s(x,y)=1$ if $x=y=0$. The cyclomatic complexity is defined as
\begin{equation}
C = E - N + 2, \qquad C' = E' - N' + 2
\end{equation}
Then, the CFG structural similarity $S_{\mathrm{cfg}}$ is
\begin{equation}
S_{\mathrm{cfg}} = \frac{1}{3} \left\{ s(N,N') + s(E,E') + s(C,C') \right\}
\end{equation}
That is, for the CFGs of the original program and the decompiled program, we compute the similarities of the number of nodes, the number of edges, and the cyclomatic complexity, and use their average as the final CFG structural similarity.

The CFG is an important representation in program analysis, serving as the basis for reachability analysis and data flow analysis.
For this reason, we assigned it the highest weight among the three dimensions, 50\%.

\subsubsection{Data Structure Restoration}
This evaluates the restoration status of arrays and structs. The score is computed with arrays weighted at 0.7 and structs at 0.3. The weight of this metric is 20\%.

\subsubsection{Score Calculation for Structural Similarity}
The overall structural similarity score $S_{\text{struct}}$ is computed by the following formula:
\begin{equation}
S_{\text{struct}} = 0.30 \times S_{\text{sig}} + 0.50 \times S_{\text{cfg}} + 0.20 \times S_{\text{data}}
\end{equation}
where $S_{\text{sig}}$ denotes function signature completeness, $S_{\text{cfg}}$ denotes CFG structural similarity, and $S_{\text{data}}$ denotes data structure restoration.

\subsection{Behavioral Similarity}
Behavioral similarity is a metric that evaluates whether the behavior of the program is equivalent to the original binary. It consists of data/logic operation correctness, API/system call correctness, and error handling completeness.

\subsubsection{Data/Logic Operation Correctness}
This detects errors in memory allocation and logical operations. It has been shown that the vocabulary and token occurrence patterns of source code strongly correlate with vulnerability density; by identifying description flaws with this metric, we confirm tendencies in vulnerability occurrence \citep{Scandariato2014}.

Table \ref{tab:bug_severity} shows the correspondence between bug types and severity. The errors to be detected are classified into four categories. Each error is assigned a severity, weighted as 1.0 for High severity and 0.5 for Medium severity. The weights are based on the technical impact of CWSS, a standard metric in the security field \citep{CWSS}. Items directly related to memory tampering or code execution are classified as High, and other items leading to resource management issues or improper state transitions are classified as Medium.

In this study, the number of errors $n$ is defined as a severity-weighted error count. That is,
\begin{equation}
n = \sum_{i=1}^{m} w_i
\end{equation}
where $m$ is the total number of detected errors and $w_i$ is the weight corresponding to the severity of each error, satisfying $w_i \in \{1.0, 0.5\}$. That is, High-severity errors are added as 1.0 and Medium-severity errors as 0.5.

\begin{table}[h]
\caption{Bug types and severity}
\label{tab:bug_severity}
\centering
\begin{tabular}{llc}
\hline
Category & Bug type & Severity \\
\hline
Memory management & Missing sizeof & High \\
Memory management & Missing NULL check & High \\
Memory management & Memory leak & Medium \\
\hline
Arithmetic/type conversion & Unit conversion error & High \\
Arithmetic/type conversion & Bitwise operation misuse & High \\
Arithmetic/type conversion & Integer overflow & Medium \\
Arithmetic/type conversion & Missing division-by-zero check & Medium \\
\hline
Pointer/array & Out-of-bounds array access & High \\
Pointer/array & Uninitialized pointer & High \\
Pointer/array & Missing NULL termination & Medium \\
\hline
Function call & Insufficient arguments & High \\
Function call & Ignored return value & Medium \\
\hline
\end{tabular}
\end{table}

The data/logic operation correctness score $S_{\text{logic}}$ is computed from the weighted total of detected errors $n$ as follows.
\begin{equation}
S_{\text{logic}} =
\begin{cases}
1.0 & (n = 0) \\
\max(0.0, 1.0 - 0.15n) & (0 < n \le 5) \\
\max(0.2, 1.0 - 0.10n) & (n > 5)
\end{cases}
\end{equation}

First, when the weighted total of errors is small ($0 < n \le 5$), the deduction per unit is set to a relatively large 0.15 so that subtle differences are reflected in the score. When $n$ exceeds 5, the deduction is relaxed to 0.10 and the lower bound of the score is set to 0.2. These errors are detected by pattern matching using regular expressions. This design reflects the fact that even when many behavioral defects exist, the operation logic and control flow still retain a certain correlation with the original. The weight of this metric is 40\%.

\subsubsection{API/System Call Correctness}
This evaluates the call order and call set of the program. API similarity and system call similarity are each integrated with a weight of 0.5. For the similarity computation, SequenceMatcher is used to evaluate the number of calls preserving order, and the Jaccard coefficient is used to evaluate set similarity.
SequenceMatcher
detects the longest common subsequence between two sequences and computes the ratio of order-preserving matches \citep{ratcliff1988pattern}. For example, if the original API call order is [malloc, open, read, close] and the decompilation result is [malloc, read, close], three order-preserving calls match, yielding a high score.

The Jaccard coefficient is the number of common elements of two sets divided by the total number of elements, and measures the similarity of API call types while ignoring order \citep{Jaccard}. For example, if the original is \{malloc, open, read, close\} and the decompilation result is \{malloc, read, close, write\}, the common elements are the three \{malloc, read, close\} and the total elements are the five \{malloc, open, read, close, write\}, so the Jaccard coefficient is 3/5 = 0.6.

The final score is computed as a weighted average of SequenceMatcher at 60\% and the Jaccard coefficient at 40\%. System call similarity is computed in the same way, and API similarity and system call similarity are integrated with weights of 0.5 each.
If recompilation fails, the score is 0, imposing a substantial penalty. The weight of this metric is 30\%.

\subsubsection{Error Handling Completeness}
Error handling completeness evaluates uninitialized structs and missing error checks.
Uninitialized structs are classified as Medium with a weight of 0.5, and missing error checks as High with a weight of 1.0.

In particular, the detection targets are missing error checks for resource operations such as \texttt{malloc}, \texttt{calloc}, \texttt{open}, and \texttt{fopen}.
Because these functions return \texttt{NULL} or \texttt{-1} on failure, omitting these checks can cause runtime errors and vulnerabilities through NULL dereference or the use of invalid file descriptors.

In this study, no standalone score is defined for missing error checks; instead, they are evaluated together with uninitialized structs as an integrated error handling completeness score $S_{\text{error}}$.
The error handling completeness score $S_{\text{error}}$ is defined, using the number of uninitialized structs $n_{\text{struct}}$ and the number of missing error checks $n_{\text{check}}$, as
\begin{equation}
S_{\text{error}} = \max\left(0.3,\ 1.0 - 0.5 n_{\text{struct}} - 1.0 n_{\text{check}}\right)
\end{equation}
That is, 0.5 is deducted per uninitialized struct and 1.0 per missing error check, with a lower bound of 0.3. This lower bound takes into account that even when error handling defects exist, the main processing logic itself may still be preserved.
The weight of this metric is 30\%.

\subsubsection{Score Calculation for Behavioral Similarity}
The overall behavioral similarity score $S_{\text{behav}}$ is computed by the following formula:

\begin{equation}
S_{\text{behav}} = 0.40 \times S_{\text{logic}} + 0.30 \times S_{\text{api}} + 0.30 \times S_{\text{error}}
\end{equation}
where $S_{\text{logic}}$ denotes data/logic operation correctness, $S_{\text{api}}$ denotes API/system call correctness, and $S_{\text{error}}$ denotes error handling completeness.

\subsection{Semantic Similarity}
Semantic similarity is a metric that evaluates the semantic quality of the program, consisting of three items: identifier usage, variable name consistency, and type system utilization.

\subsubsection{Identifier Usage}
Identifier usage evaluates the avoidance of magic numbers and the appropriate use of constants.

First, magic numbers in the source code are detected. A magic number is a numeric literal written directly in the code. Common values such as 0, 1, 2, 10, and 100 are excluded. A penalty is imposed according to the number of detected magic numbers. The penalty is computed as the count$\times$0.05, capped at 0.5. The base score is 1.0 minus the penalty.

Next, a bonus is added for the appropriate use of constants. A bonus of +0.2 is given for using \#define macro definitions and +0.1 for using const declarations. The final score is normalized to the range 0.0 to 1.0.

For example, if five magic numbers are detected and both \#define and const are used, the score is 1.0 - (5$\times$0.05) + 0.2 + 0.1 = 1.05, which is normalized to 1.0. The weight of this metric is 50\%.

\subsubsection{Variable Name Consistency}
This evaluates the degree of agreement of variable names between the original and the decompilation result.
This metric compares the variable names contained in the decompiled source code with those in the original source code. In general, original variable names are lost in decompilation from optimized binaries, so rule-based methods such as Ghidra use automatically generated variable names, resulting in low scores. In contrast, LLM-based methods may infer meaningful variable names based on their training data, and this metric evaluates that degree of restoration.

First, variable names are extracted from both source codes. Variable definitions accompanied by type declarations are used for extraction. Next, common variable names such as i, j, k, ret, tmp, n, fd, and len are excluded.

The score is computed by dividing the number of common variable names by the number of variable names in the original.
For example, consider the case where the original variable set is $V_{\text{orig}} = \{\text{count}, \text{buffer}, \text{total}\}$ and the decompiled variable set is $V_{\text{dec}} = \{\text{count}, \text{buf}, \text{total}\}$. The common variables are the two in $V_{\text{orig}} \cap V_{\text{dec}} = \{\text{count}, \text{total}\}$. The score is $|V_{\text{orig}} \cap V_{\text{dec}}| / |V_{\text{orig}}| = 2 / 3 = 0.667$.
This indicates that 66.7\% of the variable names are shared. The weight of this metric is 30\%.

\subsubsection{Type System Utilization}
Type system utilization evaluates the appropriate use of type definitions.

The base score is 0.5. A bonus of +0.2 is given for the use of typedef and +0.3 for the use of standard headers. However, a penalty of -0.1 is imposed if structs are defined without using standard headers. A bonus of +0.1 is given for the use of fixed-width integer types such as int8\_t and uint32\_t. The final score is normalized to the range 0.0 to 1.0.
The weight of this metric is 20\%.

\subsubsection{Score Calculation for Semantic Similarity}
The overall semantic similarity score $S_{\text{sem}}$ is computed by the following formula:
\begin{equation}
S_{\text{sem}} = 0.50 \times S_{\text{ident}} + 0.30 \times S_{\text{var}} + 0.20 \times S_{\text{type}}
\end{equation}
where $S_{\text{ident}}$ denotes identifier usage, $S_{\text{var}}$ denotes variable name consistency, and $S_{\text{type}}$ denotes type system utilization.

\subsection{Overall Evaluation Score}

The final overall evaluation score $S_{\text{total}}$ is computed as a weighted average of the three major categories by the following formula:
\begin{equation}
S_{\text{total}} = 0.35 \times S_{\text{struct}} + 0.40 \times S_{\text{behav}} + 0.25 \times S_{\text{sem}}
\end{equation}

This weighting takes into account that equivalence, the goal of decompilation, is primarily indicated by behavioral similarity. Therefore, behavioral similarity is given the highest weight of 0.40. Structural similarity is set to 0.35, and semantic similarity to 0.25.

\begin{table}[h]
\caption{Dataset breakdown by category}
\label{tab:dataset}
\centering
\begin{tabular}{lr}
\hline
Category & Number of programs \\
\hline
Network Services & 46 \\
System Utilities & 61 \\
Hardware/Drivers & 50 \\
Libraries/Helpers & 62 \\
Firewall/Security & 55 \\
Package Management & 44 \\
\hline
Total & 318 \\
\hline
\end{tabular}
\end{table}

\begin{figure}[h]
\centering
\includegraphics[width=\textwidth]{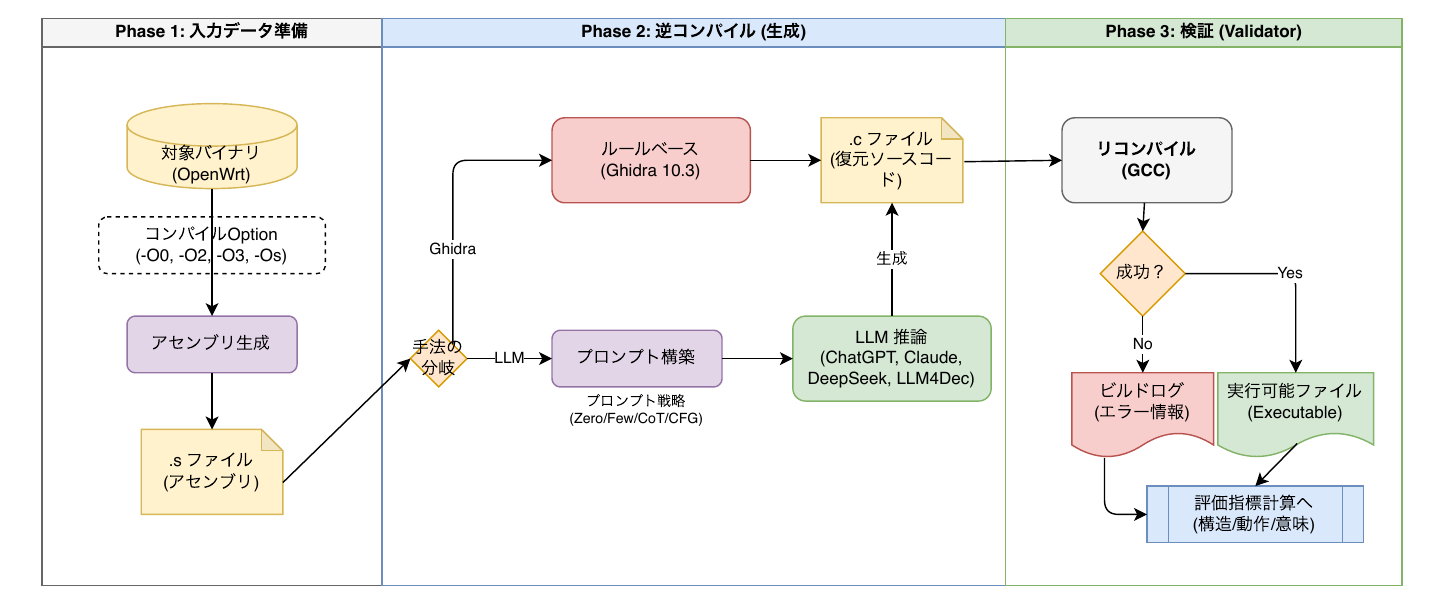}
\caption{Overview of the evaluation system}
\label{fig:systemarchtecture}
\end{figure}

\section{Experiments}
\subsection{Experimental Overview}
We decompile programs from IoT firmware and evaluate the generation results of each model based on the proposed metrics.
The target programs were taken from OpenWrt, an open-source router program. OpenWrt is a Linux-based firmware for routers \citep{openwrt_project}. It is adopted as the basis of many commercial routers, and its vulnerabilities propagate through the supply chain to an enormous number of devices, making it an ideal experimental target \citep{10.1145/3427228.3427294}.

Figure \ref{fig:systemarchtecture} shows the evaluation system. The evaluation system consists of three phases. In the first phase, binaries and assembly files are generated for each program with all compiler options.
In this experiment, 318 programs were extracted from the OpenWrt router. They were classified into six categories to avoid bias in program types. The category breakdown is shown in Table \ref{tab:dataset}.
They were compiled with GCC 11.4 for the x86-64 architecture to generate ELF-format binaries and assembly. All optimization options -O0, -O2, -O3, and -Os were used.

In the second phase, the assembly is fed into the decompilers to generate C files.
As decompilers, we used Ghidra 10.3, ChatGPT-5.1, Claude Haiku 4.5, DeepSeek-R1, and LLM4Decompile. Four prompt strategies were applied to the LLMs. Zero-shot gives only a basic decompilation instruction, and Few-shot includes three examples. Chain-of-Thought encourages step-by-step reasoning, and CFG-shot adds CFG information to the assembly.

In the third phase, the generated C files are automatically compiled to check re-executability, and the generated C programs and executables are analyzed to compute scores based on the proposed metrics.

The workstation used for the experiments is equipped with an Intel Xeon w7-2575X processor, 128GB RAM, and an NVIDIA RTX PRO 6000. The operating system is Ubuntu 22.04 LTS.





\begin{figure}[h]
\centering
\includegraphics[width=0.8\textwidth]{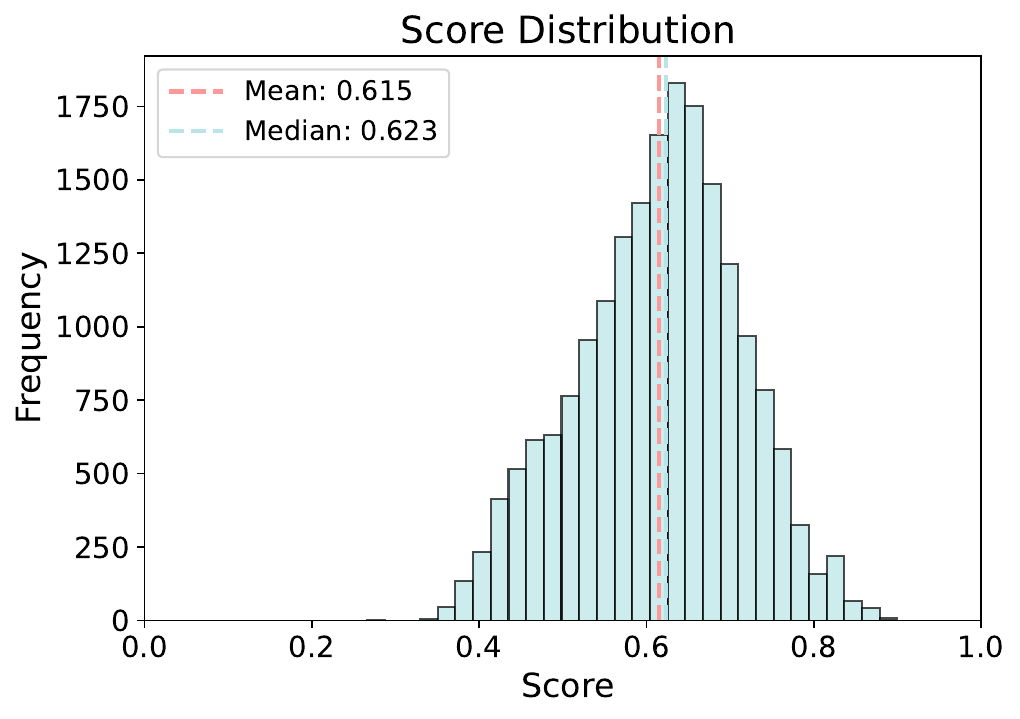}
\caption{Distribution of overall scores}
\label{fig:score_dist}
\end{figure}
\subsection{Experimental Results}
\subsubsection{Score Distribution}
When decompilation was performed on a total of 21,624 C programs,
generation errors occurred during decompilation for 2,299 programs,
and 402 programs had at least one of the nine metrics unmeasurable.
Accordingly, in this experiment we analyze the 19,223 programs for which all nine metrics are measurable.
Figure \ref{fig:score_dist} shows the distribution of overall scores. With 19,223 samples, a mean of 0.615, and a median of 0.623, the distribution is approximately normal, indicating that the dataset has appropriate variance.

\begin{table}[h]
\caption{Recompilation success rate by model}
\label{tab:recompile}
\centering
\begin{tabular}{lrrr}
\hline
Model & Success & Failure & Success rate \\
\hline
ChatGPT-5.1 & 910 & 4,104 & 18.1\% \\
Claude Haiku 4.5 & 880 & 4,088 & 17.7\% \\
DeepSeek-R1 & 906 & 3,992 & 18.5\% \\
Ghidra & 0 & 213 & 0.0\% \\
LLM4Decompile & 40 & 4,492 & 0.9\% \\
\hline
\end{tabular}
\end{table}

\subsubsection{Recompilation Success Rate}
The overall recompilation success rate was 13.9\% (2,736 out of 19,625). The success rates by model are shown in Table \ref{tab:recompile}. The total differs from the previous section because the previous section excluded cases with missing metrics.
The reason Ghidra had zero successes is that Ghidra aims at faithful rule-based restoration on a per-function basis and does not supply the necessary libraries, so no executable files were generated. In addition, LLM4Decompile, which was trained specifically for decompilation, also had a success rate of only 0.9\%, showing that dedicated training does not necessarily lead to improved performance.

\subsubsection{Comparison Between Recompilation Success and Failure Groups}
Table \ref{tab:success} shows the results of comparing the evaluation scores of the recompilation success and failure groups using t-tests.
In this study, because the sample size is extremely large, even slight differences tend to be statistically significant ($p < 0.001$). Therefore, to assess the magnitude of substantive differences, we adopt the effect size measured by Cohen's $d$ as the primary criterion \citep{sullivan2012using}.

\begin{table}[h]
\caption{Score comparison between recompilation success and failure groups}
\label{tab:success}
\centering
\begin{tabular}{lrrrl}
\hline
Metric & Success & Failure & Cohen's $d$ & Effect size \\
\hline
Overall score & 0.689 & 0.603 & 0.92 & Large \\
Structural similarity & 0.603 & 0.460 & 0.69 & Medium \\
Behavioral similarity & 0.787 & 0.728 & 0.96 & Large \\
Semantic similarity & 0.651 & 0.604 & 0.33 & Small \\
\hline
\end{tabular}
\end{table}

\begin{figure}[h]
\centering
\includegraphics[width=\textwidth]{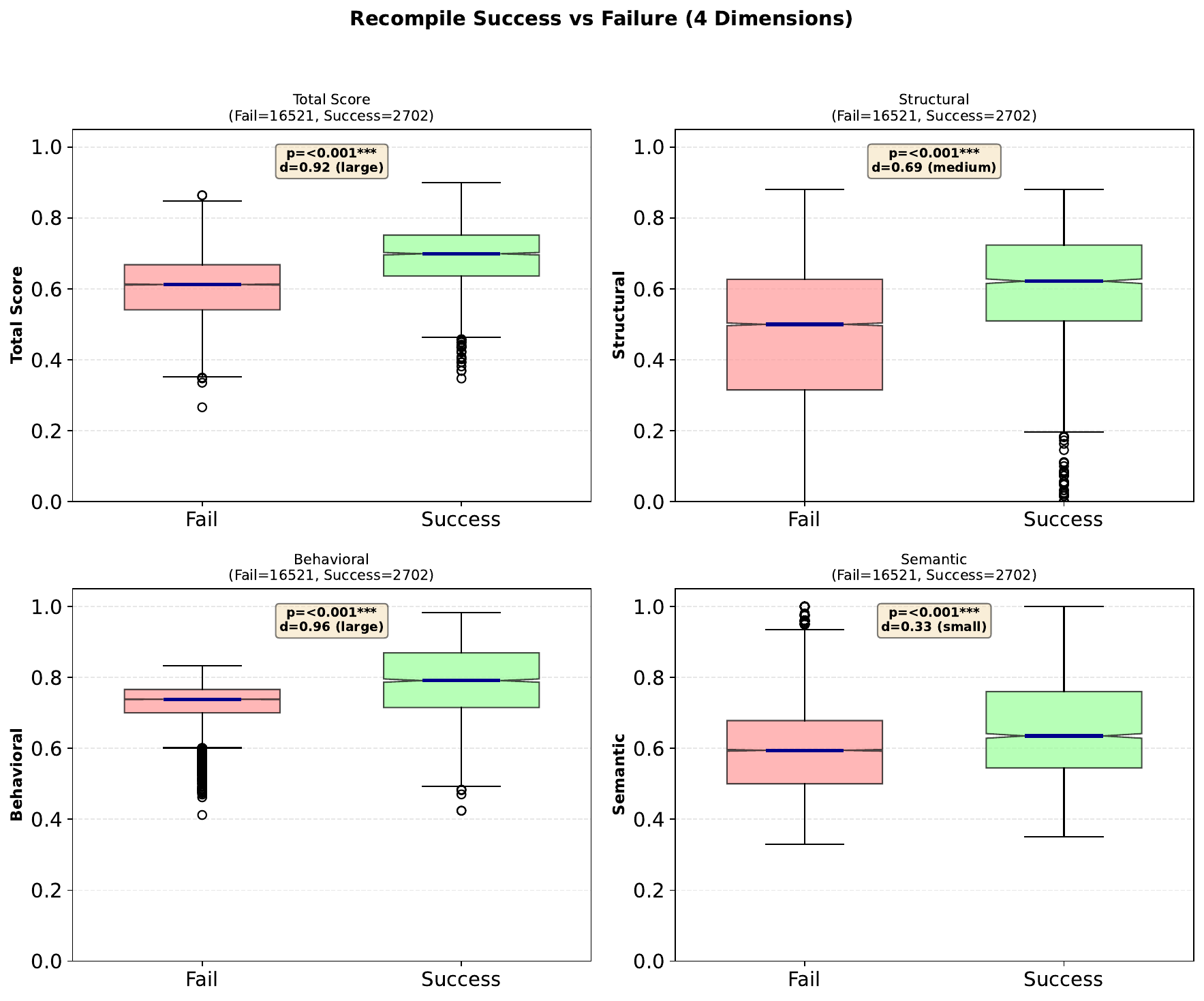}
\caption{Score comparison between recompilation success and failure groups}
\label{fig:success_comparison}
\end{figure}

Figure \ref{fig:success_comparison} shows the score comparison. The recompilation success group contains 2,736 samples and the failure group 16,889 samples. The success group had significantly higher overall scores than the failure group, with Cohen's $d=0.92$, indicating a large difference. This confirms that the proposed evaluation metrics function appropriately. Among the three major categories, behavioral similarity showed the largest difference at $d=0.96$, followed by structural similarity with a medium difference at $d=0.69$, and semantic similarity with a small difference at $d=0.33$.

\subsubsection{Analysis by Sub-metric}
\begin{figure}[h]
\centering
\includegraphics[width=0.8\textwidth]{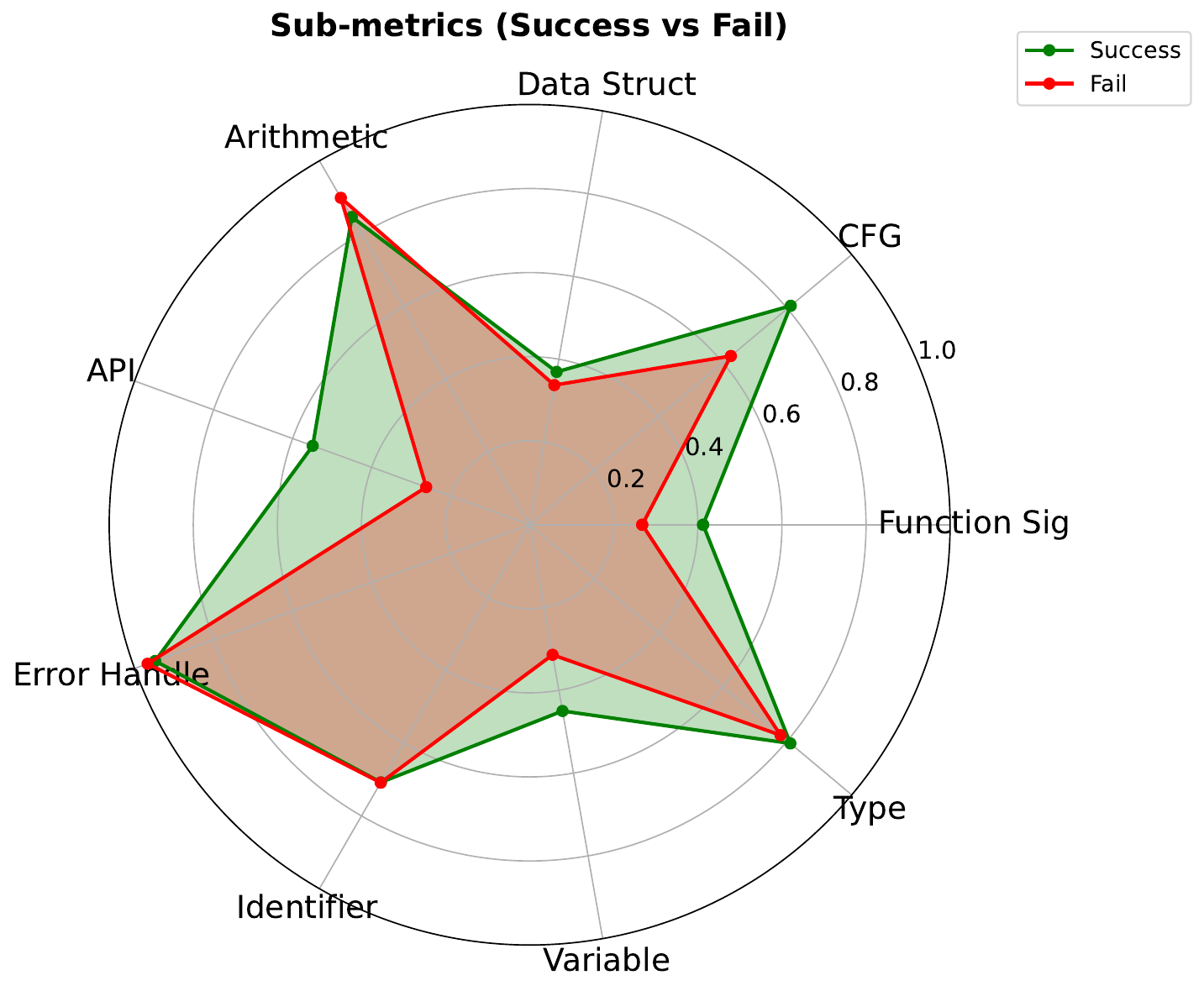}
\caption{Radar chart of sub-metrics}
\label{fig:submetrics_radar}
\end{figure}

Table \ref{tab:submetric} shows the results of comparing the recompilation success and failure groups by t-test for the nine sub-metrics. Figure \ref{fig:submetrics_radar} shows the score differences by sub-metric.

\begin{table}[h]
\caption{Score comparison by sub-metric (all $p<0.001$***)}
\label{tab:submetric}
\centering
\begin{tabular}{llrr}
\hline
Major category & Sub-metric & Cohen's $d$ & Effect size \\
\hline
Structure & Function signature & 0.65 & Medium \\
Structure & CFG structure & 0.61 & Medium \\
Structure & Data structure & 0.10 & Negligible \\
Behavior & Logic operations & $-0.49$ & Small (inverse) \\
Behavior & API detection & 1.91 & Very Large \\
Behavior & Error handling & $-0.35$ & Small (inverse) \\
Semantics & Identifiers & $-0.00$ & Negligible (inverse) \\
Semantics & Variable names & 0.36 & Small \\
Semantics & Type information & 0.18 & Negligible \\
\hline
\end{tabular}
\end{table}

API detection, a sub-metric of behavioral similarity, showed the largest effect size at $d=1.91$, revealing that accurate restoration of external library calls is decisively important for recompilation success. Next, function signature, a sub-metric of structural similarity, showed a medium difference at $d=0.65$, and CFG structure a medium difference at $d=0.61$. This indicates that behavioral and structural similarity are important for recompilation success. On the other hand, logic operations, a sub-metric of behavioral similarity, showed a negative effect size of $d=-0.49$, an unexpected result in which the recompilation success group scored lower. This suggests that there is room for improvement in the metric design.

\subsubsection{Complementary Relationship Between Behavioral and Structural Evaluation}
\begin{figure}[h]
\centering
\includegraphics[width=\textwidth]{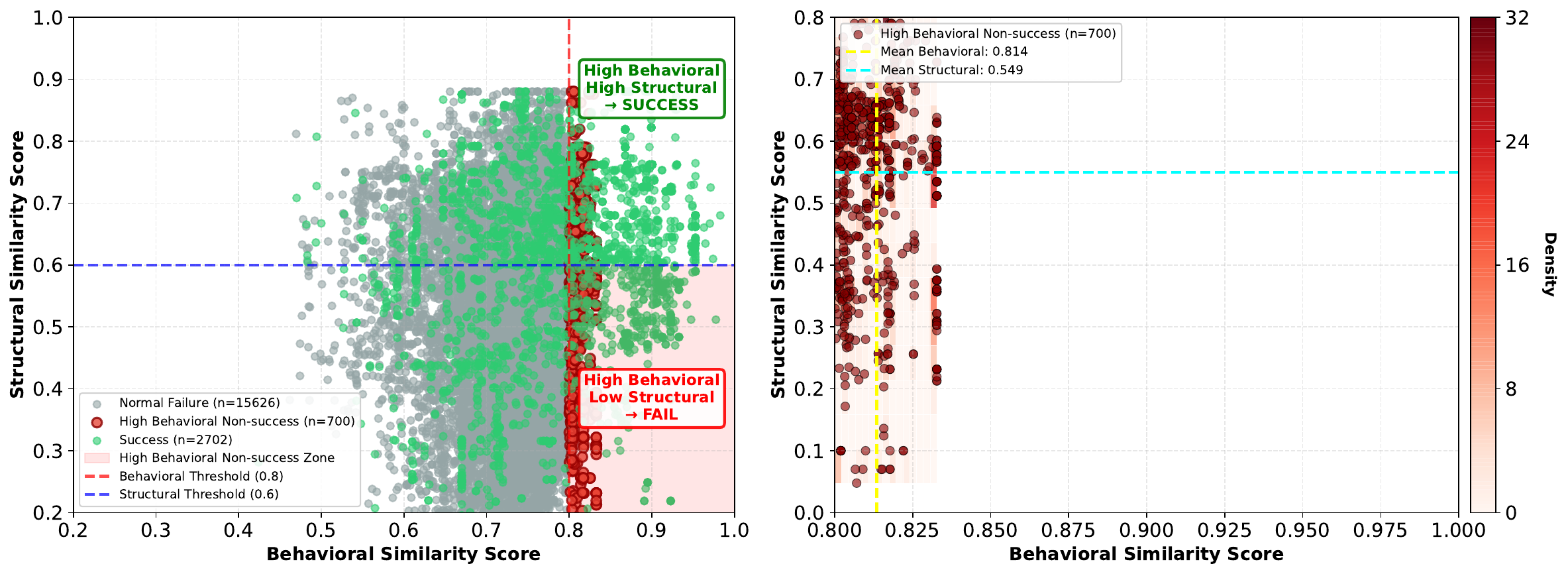}
\caption{Distribution of behavioral vs.\ structural similarity}
\label{fig:complementary}
\end{figure}

We analyzed the high-behavior failure group of 700 cases: cases in which recompilation failed despite a behavioral similarity of 0.8 or higher.
In this section, only LLM-generated results are targeted. Ghidra is not a method whose main purpose is to generate complete, recompilable source code, and failures in Ghidra differ in nature from failures stemming from the generation quality of LLMs, so it was excluded from the analysis in this section. On the other hand, because Ghidra is important for comparison with existing methods, it was retained in the model-by-model comparison and as reference values.
The analysis results are shown in Table \ref{tab:false_positive}. Figure \ref{fig:complementary} shows the distribution of behavioral and structural similarity.

\begin{table}[h]
\caption{Classification of recompilation failure cases}
\label{tab:false_positive}
\centering
\begin{tabular}{lrrrr}
\hline
Classification & Count & Ratio & Behavioral sim. & Structural sim. \\
\hline
Success & 2,702 & 14.1\% & 0.787 & 0.603 \\
High-behavior failure & 700 & 3.6\% & 0.814 & 0.549 \\
Ordinary failure & 15,821 & 82.3\% & 0.724 & 0.456 \\
\hline
Total & 19,223 & 100.0\% & 0.736 & 0.480 \\
\hline
\end{tabular}
\end{table}

The high-behavior failure group consists of 700 cases, accounting for 3.6\% of the total and 4.2\% of the 16,521 failure cases.
The behavioral similarity of the high-behavior failure group is 0.814, exceeding the success group's 0.787 and showing the highest value. However, its structural similarity remains at 0.549, below the success group's 0.603. This shows that high behavioral similarity does not necessarily guarantee recompilation success. Comparing the success group and the high-behavior failure group, the high-behavior failure group is 0.027 higher in behavioral similarity but 0.054 lower in structural score. These structural defects are the factor causing recompilation to fail for code that appears behaviorally correct.

This result shows that there exist structural defects that are difficult to detect with behavioral evaluation alone. Even if API calls and system calls are correctly restored and there are no errors in logic operations, recompilation fails if the control flow is incorrect. Specific examples include missing resource release in error branches and deficiencies in branch structures such as boundary checks. These defects cannot be detected by evaluating the presence or order of API calls, but can be detected by evaluating the CFG structure.


\subsubsection{Analysis by Compiler Option}

Table \ref{tab:optimization} shows the recompilation success rate by compiler option.

\begin{table}[h]
\caption{Recompilation success rate by compiler option}
\label{tab:optimization}
\centering
\begin{tabular}{lr}
\hline
Option & Success rate \\
\hline
-O0 & 15.3\% \\
-O2 & 12.7\% \\
-O3 & 12.6\% \\
-Os & 13.1\% \\
\hline
\end{tabular}
\end{table}

Linear regression analysis showed a negative correlation tendency between optimization level and success rate, with $r=-0.683$, $p=0.317$. While -O0 showed a success rate of 15.3\%, it dropped to 12.6\% at -O3. We attribute this to the loss of the original structure caused by optimizations such as inline expansion, loop unrolling, and instruction reordering.


\subsubsection{Analysis by Prompt Strategy}

Table \ref{tab:prompt} shows the results of comparing overall scores by prompt type using one-way analysis of variance.

\begin{table}[h]
\caption{Comparison of overall scores by prompt}
\label{tab:prompt}
\centering
\begin{tabular}{lr}
\hline
Prompt & Mean score \\
\hline
Chain-of-Thought & 0.616 \\
Default & 0.543 \\
Few-shot & 0.614 \\
CFG-shot & 0.620 \\
Zero-shot & 0.615 \\
\hline
\end{tabular}
\end{table}

A statistically significant difference was found ($F(3, 19621)=30.00$, $p<0.001$***), but the effect size was very small ($\eta^2=0.006$, Small effect). This result indicates that prompt engineering does not have a large impact on decompilation quality. In particular, even the CFG-shot prompt, which adds CFG information, showed no notable effect.
While structural similarity is a strong predictor, a possible reason why CFG-shot, which provides CFG information, did not contribute is that presenting the CFG in JSON format duplicated already-known information and may have functioned as noise that dispersed the model's attention.

\subsubsection{Analysis by Model}
Table \ref{tab:model} shows the results of comparing overall scores by model using one-way analysis of variance.

\begin{table}[h]
\caption{Comparison of overall scores by model}
\label{tab:model}
\centering
\begin{tabular}{lr}
\hline
Model & Mean score \\
\hline
DeepSeek-R1 & 0.643 \\
Claude Haiku 4.5 & 0.642 \\
ChatGPT-5.1 & 0.639 \\
Ghidra & 0.543 \\
LLM4Decompile & 0.533 \\
\hline
\end{tabular}
\end{table}

A statistically significant difference was found with $F(4, 19620)=1401.06$, $p<0.001$***, and the effect size was large at $\eta^2=0.226$. The general-purpose LLMs were all comparable with mean scores around 0.64, exceeding the rule-based method Ghidra and the decompilation-specific LLM4Decompile by about 0.1 points. This result indicates that LLM-centered research and development is currently appropriate, and that general-purpose LLMs show higher accuracy than decompilation-specific LLMs.


\subsubsection{Comparison with Existing Code Generation Metrics}
To compare the proposed metrics with existing code generation evaluation metrics, we applied CodeBLEU and CodeScore. CodeBLEU and CodeScore can be computed as long as a correspondence with the reference source code is available. On the other hand, MSSC and D-Score are evaluation metrics for intermediate artifacts and are difficult to compare under identical conditions, so this experiment was limited to CodeBLEU and CodeScore.

Table \ref{tab:baseline_comparison} shows the overall summary and the comparison between the success-group and failure-group means. CodeBLEU showed a mean of 0.265 and a median of 0.284, and a large difference of Cohen's $d=0.862$ was confirmed between the success-group mean of 0.349 and the failure-group mean of 0.252. This suggests that CodeBLEU is an auxiliary metric capable of distinguishing recompilation success from failure to some extent.

In contrast, CodeScore concentrated at high values with a mean of 0.771 and a median of 0.972, and the effect size for the difference between the success-group mean of 0.831 and the failure-group mean of 0.761 remained at Cohen's $d=0.21$. This result suggests that CodeScore cannot sufficiently discriminate quality differences in the setting of this experiment.


From the above, this experiment suggests that CodeBLEU is useful as a metric for measuring the similarity between generated code and the original code, but it is difficult for it to evaluate in detail where the code is similar. In contrast, this study also targets structural aspects such as the restoration of error handling and branch structures for quality evaluation, and is characterized by its ability to evaluate similarity from multiple perspectives.

\begin{table}[h]
\caption{Comparison of existing code generation metrics}
\label{tab:baseline_comparison}
\centering
\small
\begin{tabular}{lrrrrrr}
\hline
Metric & Mean & Median & Success mean & Failure mean & Cohen's $d$ & Effect size \\
\hline
CodeBLEU  & 0.265 & 0.284 & 0.349 & 0.252 & 0.862 & Large \\
CodeScore & 0.771 & 0.972 & 0.831 & 0.761 & 0.210 & Small \\
\hline
\end{tabular}
\end{table}


\subsubsection{Analysis by Program Category}
Comparing overall scores by program category using one-way analysis of variance yielded $F(5, 19619)=73.79$, $p<0.001$***, a statistically significant difference, but the effect size was small at $\eta^2=0.023$. This indicates that quality differences across program categories are small and that the proposed evaluation metrics are applicable without bias toward particular program types.


\subsection{Limitations and Future Work}
This study has several limitations. First, although the weights of the three major categories and the sub-metrics were determined based on prior work and preliminary experiments, the optimal weights remain unclear. In the future, it will be necessary to examine equal weighting and weight settings based on principal component analysis or effect sizes.

Second, the fact that the logic operation metric showed a negative correlation suggests that there is room for improvement in the metric design. Given that the recompilation success group scored lower, the current regular-expression-based bug detection patterns may be producing false positives on complex code. In the future, we will consider migrating to context-aware syntax-tree-based analysis.

Third, this study targeted 318 programs, but validation on larger datasets is desirable. Applicability to IoT firmware other than OpenWrt and to different architectures should also be examined.


Building on the importance of API detection and CFG structure revealed in this study, we will work on the design and evaluation of decompilation methods that strengthen these aspects.

\section{Conclusion}

In this study, we proposed quality evaluation metrics for decompilation results of IoT device firmware. The proposed metric consists of three major categories---behavioral similarity, structural similarity, and semantic similarity---each with three sub-metrics, for a total of nine metrics. In our experiments, targeting 318 programs from the OpenWrt router, we statistically evaluated the quality of 19,625 results generated by five decompilation models using the proposed metrics.

The experiments revealed the following four findings. First, LLM-based decompilation methods showed higher scores than the rule-based method Ghidra, with a large effect size of $\eta^2=0.226$. This demonstrates the validity of research and development on LLM-based decompilation methods. Second, differences due to prompt strategies were statistically significant, but the effect size was small at $\eta^2=0.006$, indicating that their substantive impact is limited. Third, a negative correlation tendency was observed between optimization level and recompilation success rate, with success rates of 15.3\% at -O0 and 12.6\% at -O3. Fourth, behavioral and structural similarity were found to correlate with recompilation success. Behavioral similarity showed an effect size of Cohen's $d=0.96$ and structural similarity $d=0.69$, revealing that these are important metrics for predicting decompilation quality.
The analysis of the high-behavior failure group showed that even behaviorally correct code fails to recompile when its structural similarity is low, suggesting the importance of structural quality such as the restoration of branches.

In vulnerability analysis of IoT devices, faithfully restoring the original program is extremely important. For the goal of faithfully reproducing the original implementation, the most effective existing approach was decompilation using general-purpose LLMs. 
Furthermore, it was confirmed that behavioral similarity and structural similarity contribute greatly to recompilation success or failure. In particular, structural evaluation was effective in identifying cases that are valid at the behavioral level but contain defects in their internal structure. From these perspectives, the approach using general-purpose LLMs is a method capable of more faithfully reproducing the original implementation in terms of both behavior and structure, and is suggested to contribute to the advancement of vulnerability analysis technology.
In future work, we will focus on optimizing the metric definitions and weight settings, as well as designing and evaluating decompilation methods that emphasize the metrics with large effect sizes.

\section*{Acknowledgements}
This work was supported by JSPS KAKENHI Grant Number 25K21182.

\bibliographystyle{plainnat}
\bibliography{myrefs_en}

\begin{thebibliography}{28}
\providecommand{\natexlab}[1]{#1}
\providecommand{\url}[1]{\texttt{#1}}
\expandafter\ifx\csname urlstyle\endcsname\relax
  \providecommand{\doi}[1]{doi: #1}\else
  \providecommand{\doi}{doi: \begingroup \urlstyle{rm}\Url}\fi

\bibitem[Allen(1970)]{cfg}
Frances~E. Allen.
\newblock Control flow analysis.
\newblock In \emph{Proceedings of a Symposium on Compiler Optimization}, page
  1–19, New York, NY, USA, 1970. Association for Computing Machinery.
\newblock ISBN 9781450373869.
\newblock URL \url{https://doi.org/10.1145/800028.808479}.

\bibitem[Armengol-Estap\'{e} et~al.(2022)Armengol-Estap\'{e}, Woodruff,
  Brauckmann, Magalh\~{a}es, and O'Boyle]{exebench}
Jordi Armengol-Estap\'{e}, Jackson Woodruff, Alexander Brauckmann, Jos\'{e}
  Wesley de~Souza Magalh\~{a}es, and Michael F.~P. O'Boyle.
\newblock Exebench: an ml-scale dataset of executable c functions.
\newblock In \emph{Proceedings of the 6th ACM SIGPLAN International Symposium
  on Machine Programming}, page 50–59, New York, NY, USA, 2022. Association
  for Computing Machinery.
\newblock ISBN 9781450392730.
\newblock URL \url{https://doi.org/10.1145/3520312.3534867}.

\bibitem[Armengol-Estap{\'e} et~al.(2024)Armengol-Estap{\'e}, Woodruff,
  Cummins, and O'Boyle]{SLaDe}
Jordi Armengol-Estap{\'e}, Jackson Woodruff, Chris Cummins, and Michael F.~P.
  O'Boyle.
\newblock {SLaDe}: A portable small language model decompiler for optimized
  assembly.
\newblock In \emph{Proceedings of the 2024 IEEE/ACM International Symposium on
  Code Generation and Optimization}, pages 67--80, 2024.
\newblock URL \url{https://doi.org/10.1109/CGO57630.2024.10444788}.

\bibitem[Cifuentes and Gough(1995)]{Cifuentes1995}
Cristina Cifuentes and K.~John Gough.
\newblock Decompilation of binary programs.
\newblock \emph{Software: Practice and Experience}, 25\penalty0 (7):\penalty0
  811--829, 1995.
\newblock URL \url{https://doi.org/10.1002/spe.4380250703}.

\bibitem[Cummins et~al.(2025)Cummins, Seeker, Grubisic, Rozi{\`e}re, Gehring,
  Synnaeve, and Leather]{metallmcompiler2024}
Chris Cummins, Volker Seeker, Dejan Grubisic, Baptiste Rozi{\`e}re, Jonas
  Gehring, Gabriel Synnaeve, and Hugh Leather.
\newblock {LLM Compiler}: Foundation language models for compiler optimization.
\newblock In \emph{Proceedings of the 2025 International Symposium on Code
  Generation and Optimization}, pages 141--153, 2025.
\newblock URL \url{https://doi.org/10.1145/3708493.3712691}.

\bibitem[Dong et~al.(2025)Dong, Ding, Jiang, Li, Li, and Jin]{CodeScore2024}
Yihong Dong, Jiazheng Ding, Xue Jiang, Ge~Li, Zhuo Li, and Zhi Jin.
\newblock Codescore: Evaluating code generation by learning code execution.
\newblock \emph{ACM Transactions on Software Engineering and Methodology},
  34\penalty0 (3), 2025.
\newblock ISSN 1049-331X.
\newblock URL \url{https://doi.org/10.1145/3695991}.

\bibitem[Evtikhiev et~al.(2023)Evtikhiev, Bogomolov, Sokolov, and
  Bryksin]{Evtikhiev2023}
Mikhail Evtikhiev, Egor Bogomolov, Yaroslav Sokolov, and Timofey Bryksin.
\newblock Out of the bleu: How should we assess quality of the code generation
  models?
\newblock \emph{Journal of Systems and Software}, 203\penalty0 (C), 2023.
\newblock ISSN 0164-1212.
\newblock URL \url{https://doi.org/10.1016/j.jss.2023.111741}.

\bibitem[Feng et~al.(2023)Feng, Zhu, Han, Zhou, Wen, and
  Xiang]{Feng2023IoTSurvey}
Xiaotao Feng, Xiaogang Zhu, Qing-Long Han, Wei Zhou, Sheng Wen, and Yang Xiang.
\newblock Detecting vulnerability on iot device firmware: A survey.
\newblock \emph{IEEE/CAA Journal of Automatica Sinica}, 10\penalty0
  (1):\penalty0 25--41, 2023.
\newblock URL \url{https://doi.org/10.1109/JAS.2022.105860}.

\bibitem[Gao et~al.(2025)]{DecompileBench2025}
Zuchen Gao et~al.
\newblock {DecompileBench}: A comprehensive benchmark for evaluating
  decompilers in real-world scenarios, 2025.
\newblock URL \url{https://arxiv.org/abs/2505.11340}.
\newblock Last accessed January 6, 2026.

\bibitem[{Hex-Rays SA}(2024)]{IDAPro}
{Hex-Rays SA}.
\newblock {IDA Pro}.
\newblock \url{https://hex-rays.com/ida-pro/}, 2024.
\newblock Last accessed January 6, 2026.

\bibitem[Hu et~al.(2024)Hu, Liang, and Chen]{wong2023decgpt}
Peiwei Hu, Ruigang Liang, and Kai Chen.
\newblock {DeGPT}: Optimizing decompiler output with {LLM}.
\newblock In \emph{Network and Distributed System Security Symposium}, 2024.
\newblock URL
  \url{https://www.ndss-symposium.org/ndss-paper/degpt-optimizing-decompiler-output-with-llm/}.
\newblock Last accessed January 6, 2026.

\bibitem[Jaccard(1912)]{Jaccard}
Paul Jaccard.
\newblock The distribution of the flora in the alpine zone.
\newblock \emph{The New Phytologist}, 11\penalty0 (2):\penalty0 37--50, 1912.
\newblock ISSN 0028646X, 14698137.
\newblock URL \url{http://www.jstor.org/stable/2427226}.

\bibitem[Kim et~al.(2020)Kim, Kim, Kim, Kim, Jang, and
  Kim]{10.1145/3427228.3427294}
Mingeun Kim, Dongkwan Kim, Eunsoo Kim, Suryeon Kim, Yeongjin Jang, and Yongdae
  Kim.
\newblock Firmae: Towards large-scale emulation of iot firmware for dynamic
  analysis.
\newblock In \emph{Proceedings of the 36th Annual Computer Security
  Applications Conference}, page 733–745, New York, NY, USA, 2020.
  Association for Computing Machinery.
\newblock ISBN 9781450388580.
\newblock URL \url{https://doi.org/10.1145/3427228.3427294}.

\bibitem[McCabe(1976)]{mccabe1976complexity}
Thomas~J. McCabe.
\newblock A complexity measure.
\newblock \emph{IEEE Transactions on Software Engineering}, SE-2:\penalty0
  308--320, 1976.
\newblock URL \url{https://doi.org/10.1109/TSE.1976.233837}.

\bibitem[{Ministry of Internal Affairs and Communications}(2025)]{soumu2025}
{Ministry of Internal Affairs and Communications}.
\newblock White paper on information and communications in japan, 2025 edition.
\newblock
  \url{https://www.soumu.go.jp/johotsusintokei/whitepaper/ja/r07/html/datashu.html},
  2025.
\newblock Last accessed January 6, 2026.

\bibitem[{National Security Agency}(2019)]{NSA2019}
{National Security Agency}.
\newblock Ghidra software reverse engineering framework.
\newblock \url{https://ghidra-sre.org/}, 2019.
\newblock Last accessed January 6, 2026.

\bibitem[{OpenWrt Project}(2004)]{openwrt_project}
{OpenWrt Project}.
\newblock Open{W}rt {W}ireless {F}reedom.
\newblock \url{https://openwrt.org/}, 2004.
\newblock Last accessed January 6, 2026.

\bibitem[Paul et~al.(2024)Paul, Zhu, and Bayley]{BenchmarksMetrics2024}
Debalina~Ghosh Paul, Hong Zhu, and Ian Bayley.
\newblock Benchmarks and metrics for evaluations of code generation: A critical
  review, 2024.
\newblock URL \url{https://doi.org/10.1109/AITest62860.2024.00019}.

\bibitem[Ratcliff and Metzener(1988)]{ratcliff1988pattern}
John~W. Ratcliff and David~E. Metzener.
\newblock Pattern matching: The gestalt approach.
\newblock \emph{Dr. Dobb's Journal}, 13\penalty0 (7):\penalty0 46--51, 1988.
\newblock URL \url{https://doi.org/10.4300/JGME-D-12-00156.1}.

\bibitem[Ren et~al.(2020)Ren, Guo, Lu, Zhou, Liu, Tang, Sundaresan, Zhou,
  Blanco, and Ma]{CodeBLEU2020}
Shuo Ren, Daya Guo, Shuai Lu, Long Zhou, Shujie Liu, Duyu Tang, Neel
  Sundaresan, Ming Zhou, Ambrosio Blanco, and Shuai Ma.
\newblock {CodeBLEU}: a method for automatic evaluation of code synthesis,
  2020.
\newblock URL \url{https://arxiv.org/abs/2009.10297}.
\newblock Last accessed January 6, 2026.

\bibitem[Scandariato et~al.(2014)Scandariato, Walden, Hovsepyan, and
  Joosen]{Scandariato2014}
Riccardo Scandariato, James Walden, Aram Hovsepyan, and Wouter Joosen.
\newblock Predicting vulnerable software components via text mining.
\newblock \emph{IEEE Transactions on Software Engineering}, 40\penalty0
  (10):\penalty0 993--1006, 2014.
\newblock URL \url{https://doi.org/10.1109/TSE.2014.2340398}.

\bibitem[Sirlanci et~al.(2025)Sirlanci, Yagemann, and Lin]{Sirlanci2025}
Melih Sirlanci, Carter Yagemann, and Zhiqiang Lin.
\newblock An empirical study of c decompilers: Performance metrics and error
  taxonomy.
\newblock In \emph{Proceedings of the 20th ACM Asia Conference on Computer and
  Communications Security}, page 1707–1723, New York, NY, USA, 2025.
  Association for Computing Machinery.
\newblock ISBN 9798400714108.
\newblock URL \url{https://doi.org/10.1145/3708821.3733877}.

\bibitem[Sullivan and Feinn(2012)]{sullivan2012using}
Gail Sullivan and Richard Feinn.
\newblock Using effect size—or why the p value is not enough.
\newblock \emph{Journal of graduate medical education}, 4:\penalty0 279--82,
  2012.
\newblock URL \url{https://doi.org/10.4300/JGME-D-12-00156.1}.

\bibitem[Tan et~al.(2024)Tan, Luo, Li, and Zhang]{Tan2024}
Hanzhuo Tan, Qi~Luo, Jing Li, and Yuqun Zhang.
\newblock {LLM4Decompile}: Decompiling binary code with large language models.
\newblock In \emph{Proceedings of the 2024 Conference on Empirical Methods in
  Natural Language Processing}, 2024.
\newblock URL \url{https://doi.org/10.18653/v1/2024.emnlp-main.203}.

\bibitem[{The MITRE Corporation}(2014)]{CWSS}
{The MITRE Corporation}.
\newblock Common weakness scoring system (cwss).
\newblock \url{https://cwe.mitre.org/cwss/}, 2014.
\newblock Last accessed January 6, 2026.

\bibitem[Wang and Shoshitaishvili(2017)]{angr}
Fish Wang and Yan Shoshitaishvili.
\newblock Angr -- the next generation of binary analysis.
\newblock In \emph{Proceedings of 2017 IEEE Cybersecurity Development}, pages
  8--9, 2017.
\newblock URL \url{https://doi.org/10.1109/SecDev.2017.14}.

\bibitem[Wei et~al.(2025)Wei, Wei, Geng, and Yang]{IoTSecuritySurvey2024}
Zihan Wei, Qiang Wei, Yangyang Geng, and Yahui Yang.
\newblock A survey on iot security: Vulnerability detection and protection.
\newblock In \emph{Proceedings of the 2024 International Conference on
  Artificial Intelligence of Things and Computing}, page 1–8, New York, NY,
  USA, 2025. Association for Computing Machinery.
\newblock ISBN 9798400709869.
\newblock URL \url{https://doi.org/10.1145/3708282.3708283}.

\bibitem[Zou et~al.(2025)Zou, Khan, Wu, Gao, Bianchi, and Tian]{DLiFT2025}
Muqi Zou, Arslan Khan, Ruoyu Wu, Han Gao, Antonio Bianchi, and Dave~Jing Tian.
\newblock {D-LiFT}: Improving {LLM}-based decompiler backend via code
  quality-driven fine-tuning, 2025.
\newblock URL \url{https://arxiv.org/abs/2506.10125}.
\newblock Last accessed January 6, 2026.

\end{thebibliography}

\end{document}